\documentclass[preprintnumbers,prd,showpacs,floatfix,superscriptaddress,nofootinbib,twocolumn]{revtex4-1}
\pdfoutput=1
\usepackage{graphicx,epsfig}
\usepackage{amsmath}
\usepackage {amssymb}
\usepackage {longtable}
\usepackage{multirow}
\usepackage{dcolumn}
\usepackage{bm}
\usepackage{amsfonts}
\usepackage{subfigure}
\usepackage{color}
\usepackage{relsize}
\usepackage{tikz}
\usepackage{float}

\begin{document}

\title{Existence and stability of relativistic fluid spheres supported by thin shells}

\author{Jo\~{a}o Lu\'{i}s Rosa}
\email{joaoluis92@gmail.com}
\affiliation{Centro Multidisciplinar de Astrofisica - CENTRA,
Instituto Superior T\'{e}cnico - IST,
Universidade de Lisboa - UL,
Avenida Rovisco Pais 1, 1049-001, Portugal}

\author{Pedro Pi\c{c}arra}
\email{pedro.picarra@tecnico.ulisboa.pt}
\affiliation{Centro Multidisciplinar de Astrofisica - CENTRA,
Instituto Superior T\'{e}cnico - IST,
Universidade de Lisboa - UL,
Avenida Rovisco Pais 1, 1049-001, Portugal}

\date{\today}

\begin{abstract} 
We propose two models for constant density relativistic perfect-fluid spheres supported by thin shell configurations. These models are obtained from the Schwarzschild constant density star solution: the first via the collapse of the external layers of the fluid into a thin shell by performing a matching with the exterior Schwarzschild solution at a matching radius smaller than the star radius; and the second via the creation of a vacuum bubble inside the star by matching it with an interior Minkowski spacetime. Both models are shown to satisfy both the weak and the strong energy conditions (WEC and SEC) and can have a compactness arbitrarily close to that of a black-hole without developing singularities at the center, thus being exceptions to the Buchdahl limit. We compute the stability regimes of the models proposed and we show that there are combinations of the star radius $R$ and the matching radius $R_\Sigma$ for which the solutions are stable, the dominant energy condition (DEC) is satisfied, and the radius of the object is smaller than $3M$, implying that these models could be used as models for dark matter or exotic compact objects.
\end{abstract}

\pacs{04.50.Kd,04.20.Cv,}

\maketitle

\section{Introduction}\label{sec:intro}

In an attempt to derive solutions to the Einstein's field equations (EFE) in general relativity (GR), one often encounters a situation where a hypersurface separates the whole spacetime into two regions described by two different metric tensors, often expressed in terms of different coordinate systems. In such a case, it is natural to ask which conditions the two metric tensors must satisfy in order for the two regions to be matched smoothly at the separation hypersurface. These are called the \textit{junction conditions}.

In the context of GR, the junction conditions have been deduced long ago \cite{darmois,Israel:1966rt}. These imply that both the induced metric across the separation hypersurface and the extrinsic curvature must be continuous. Various solutions to the EFE have been obtained in this formalism, namely the Schwarzschild constant density fluid star \cite{tolman1}, the Oppenheimer-Snyder stellar collapse \cite{oppenheimer}, and the matching between Friedmann-Lemaître-Robertson-Walker spacetimes with Vaidya (and consequently, Schwarzschild) exteriors \cite{senovilla1}.

If the extrinsic curvature is discontinuous across the separation hypersurface, it is still possible to perform a matching between the two spacetime regions. However, this matching is no longer smooth: a thin shell of matter arises at the junction radius \cite{Israel:1966rt,lanczos1,lanczos2}. The thermodynamics of thin shells has been studied \cite{Martinez:1996ni,santiago}, having the shell's entropy been computed in diverse scenarios e.g. rotating shells \cite{Lemos:2017mci,Lemos:2017aol} and electrically charged shells\cite{Lemos:2015ama,Lemos:2016pyc}. Collisions of shells have also been studied numerically \cite{brito}. Alternatives to the Darmois and Israel methodology have also been proposed in this context \cite{olea}. 

In what concerns relativistic fluid spheres, a few well-known results have been established. The Buchdahl theorem \cite{buchdahl} states that if the radius of a constant density relativistic fluid sphere is smaller than a factor $9/4$ of its mass in geometrical units $G=c=1$, a few problems arise. In particular, for the Schwarzschild fluid star, this implies a divergence in the central pressure and a coordinate singularity in the $g_{tt}$ component of the metric. The so-called Buchdahl limit thus imposes a restriction for physically relevant solutions. However, Buchdahl's limit is based on a few assumptions, namely that the energy density of the star is a constant, and that the fluid is isotropic. Less restrictive bounds on the radius have been obtained in situations with fewer assumptions, e.g. for anisotropic fluid models \cite{rago1,dev1,andreasson1}, keeping the energy conditions under control. 

Indeed, the energy conditions (see \cite{curiel1} for a brief review), are important indicators of the physical relevance of relativistic fluid configurations. In particular, as seen from an observer moving in along a timelike vector field, the weak energy condition (WEC) imposes that the average energy density must be non-negative, whereas the strong energy condition (SEC) stipulates that the trace of the tidal tensor must be non-negative. Although there are well-known situations where the energy conditions are violated e.g. the observable effects of dark energy at cosmological scales \cite{farnes1,visser1}, satisfying the WEC and the SEC are important steps toward the acceptance of models for astrophysical objects.

Another important feature of astrophysically relevant spacetimes is their stability, at least in cosmological timescales. Black-hole spacetimes surrounded by thin shell configurations were shown to present stability regimes \cite{frauendiener1,brady1,alestas1} under radial perturbations, see \cite{crawford,garcia1} for a detailed description of the method. Although inconclusive, these works predict a correlation between the stability of thin shell configurations and the validity of the dominant energy condition (DEC), which allows one to predict possible stability regimes of the models proposed.

The importance of these models stands on the increasing interest in exotic compact objects, also known as ECOs. If a given model for a compact object features a surface radius smaller than the light-ring radius $r=3M$, the object will present a shadow and thus be indistinguishable from a black-hole as seen from an exterior observer, with the advantage of not presenting any singularities in the spacetime. Being poorly understood features at a fundamental level, singularities are an important problem of black-hole spacetimes and various models for black-hole mimickers have been proposed \cite{vitorpani}, which can be constrained experimentally with observations from the gravitational wave  signal \cite{franzin1,andrea1,vitor3,vitor2}.

This paper is organized as follows: in Sec.\ref{sec:review}, we review a few well-known results for the Schwarzschild fluid star, the thin shell formalism, and the energy conditions, which will be needed in the following sections; in Sec.\ref{sec:model1} we derive the first model by collapsing the outer-layers of the Schwarzschild interior solution into a thin shell at a junction radius smaller than the initial radius of the star and analyze the energy conditions; in Sec. \ref{sec:model2} we derive the second model by creating a spherical Minkowski vacuum inside the Schwarzschild interior solution and again we analyze the energy conditions; in Sec.\ref{sec:stability} we compute the stability regimes of the previous models and verify the validity of the DEC; and in Sec. \ref{sec:concl} we conclude.

\section{Framework and review}\label{sec:review}

In this section we briefly review a few concepts needed throughout the paper, namely the Schwarzschild interior solution for a fluid sphere and the problems arising from the violation of the Buchdahl limit, the junction conditions in GR and consequent thin shell formalism, and  the energy conditions to be considered, more specifically the WEC, the SEC, and the DEC.

\subsection{The Schwarzschild fluid star}\label{sub:star}

Let us consider a nonrotating and spherically symmetric sphere of incompressible relativistic perfect fluid in the context of GR. In the usual spherical coordinates $\left(t,r,\theta,\phi\right)$, the line element describing the interior of such an object is
\begin{eqnarray}\label{metricint}
ds^2=-\frac{1}{4}\left(3\sqrt{1-\frac{2M}{R}}-\sqrt{1-\frac{2r^2M}{R^3}}\right)^2dt^2+\\
+\left(1-\frac{2r^2M}{R^3}\right)^{-1}dr^2+r^2\left(d\theta^2+\sin^2\theta d\phi^2\right),\nonumber
\end{eqnarray}
for $r<R$, where $M$ is the total mass of the object and $R$ is the radius of the object. On the other hand, the exterior of this object is well described by the Schwarzschild metric, i.e.,
\begin{eqnarray}\label{metricext}
ds^2=&&-\left(1-\frac{2M}{r}\right)dt^2+\left(1-\frac{2M}{r}\right)^{-1}dr^2+,\\
&&+r^2\left(d\theta^2+\sin^2\theta d\phi^2\right),\nonumber
\end{eqnarray}
for $r>R$. As the line elements provided in Eqs.\eqref{metricint} and \eqref{metricext} are given in the same coordinates and in the limit $r\to R$ both the metrics and their respective Lie derivatives are the same, the junction between the two spacetime regions at the hypersurface $r=R$ is smooth. 

In the interior region, matter is described by an isotropic perfect fluid, i.e., the stress-energy tensor $T_a^b$ is diagonal and can be written in the form
\begin{equation}\label{stress}
T_a^b=\text{diag}\left(-\rho,p,p,p\right),
\end{equation}
where $\rho$ is the energy density, which is a constant since we assumed the fluid to be incompressible, and $p$ is the pressure, which is a function of the radial coordinate $r$ as
\begin{equation}\label{pressure}
p\left(r\right)=\rho\ \frac{\sqrt{1-\frac{2r^2M}{R^3}}-\sqrt{1-\frac{2M}{R}}}{3\sqrt{1-\frac{2M}{R}}-\sqrt{1-\frac{2r^2M}{R^3}}}.
\end{equation}
At the surface of the object $r=R$ one verifies that $p\left(R\right)=0$, as expected since the exterior solution is vacuum. On the other hand, the central pressure $p_c$ can be written in terms of the total mass $M$ and the radius of the object $R$ as
\begin{equation}\label{centralp}
p\left(0\right)\equiv p_c=\rho\ \frac{1-\sqrt{1-\frac{2M}{R}}}{3\sqrt{1-\frac{2M}{R}}-1}.
\end{equation}

At this point, the consequences of setting a radius below the Buchdahl limit, $R<9M/4\equiv R_b$ are visible. Take for example the particular case $R=R_b$. From Eq. \eqref{metricint}, we verify that $g_{tt}=0$ at the origin, which corresponds to a coordinate singularity. Furthermore, from Eq.\eqref{centralp}, we confirm that the central pressure $p_c$ diverges as $R\to R_b$, which implies that the coordinate singularity in $g_{tt}$ actually corresponds to a curvature singularity in the Ricci scalar $R$. Note also that since the denominator of the pressure in Eq.\eqref{pressure} is the same quantity inside the square of $g_{tt}$ in Eq.\eqref{metricint}, the pressure divergence and the coordinate singularity will always occur at the same radii, and so we shall refer to this feature simply as "singularity." A further decrease in the radius to the interval $2M<R<R_b$ will move the singularity outward from the center of the sphere.  We are not interested in the region $R<2M$ as it is known that in this case the object collapses into a black hole.

\subsection{Junction conditions and thin shells}\label{sub:shell}

\begin{figure}
\centering
\includegraphics[scale=0.4]{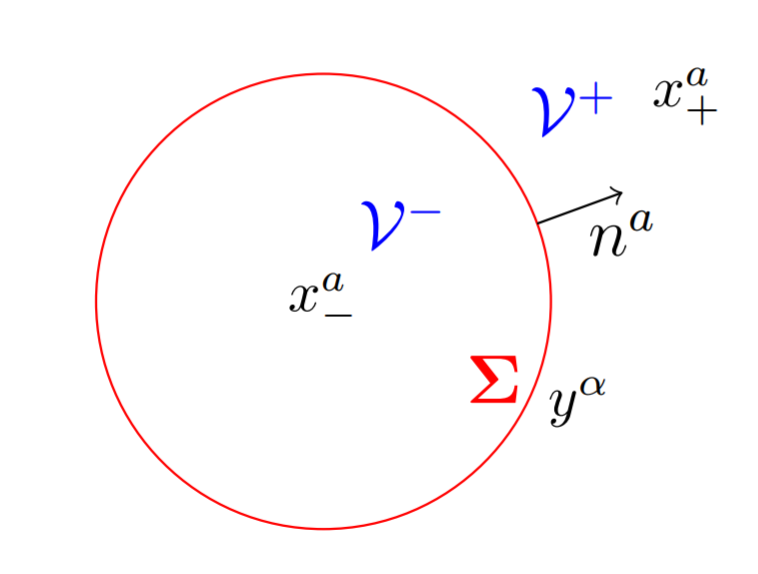}
\caption{Schematic Representation of the spacetime $\mathcal V$ divided into two regions $\mathcal V^\pm$ by a spherical hypersurface $\Sigma$ in red. $x^a_\pm$ denote the coordinate systems defined in $\mathcal V^\pm$, respectively, $y^\alpha$ denotes the coordinate system defined on $\Sigma$, and $n^a$ is the spacelike unit vector normal to $\Sigma$. }
\label{fig:junction}
\end{figure}

Let $\Sigma$ be a spacelike hypersurface that separates the spacetime $\mathcal V$ into two regions, $\mathcal V^+$ and $\mathcal V^-$. Let us consider that the metric $g_{ab}^+$, expressed in coordinates $x^a_+$, is the metric in region $\mathcal V^+$ and the metric $g_{ab}^-$, expressed in coordinates $x^a_-$, is the metric in region $\mathcal V^-$, where the latin indexes run from $0$ to $3$. Let us assume that a set of coordinates $y^\alpha$ can be defined in both sides of $\Sigma$, where greek indexes run from $0$ to $2$. See Fig.\ref{fig:junction} for a schematic representation.

The projection vectors from the 4-dimensional regions $\mathcal V^\pm$ to the 3-dimensional hypersurface $\Sigma$ are $e^a_\alpha=\partial x^a/\partial y^\alpha$. We define $n^a$ to be the unit spacelike normal vector on $\Sigma$ pointing in the direction from $\mathcal V^-$ to $\mathcal V^+$. Let $l$ denote the proper distance or time along the geodesics perpendicular to $\Sigma$ and choose $l$ to be zero at $\Sigma$, negative in the region $\mathcal V^-$, and positive in the region $\mathcal V^+$. The displacement from $\Sigma$ along the geodesics parametrized by $l$ is $dx^a=n^adl$, and $n_a=\partial_a l$. The metric $g_{ab}$ of the whole spacetime can then be written as
\begin{equation}
g_{ab}=g_{ab}^+\Theta\left(l\right)+g_{ab}^-\Theta\left(-l\right),
\end{equation}
where $\Theta\left(l\right)$ is the Heaviside distribution function, whose derivative is given by the Dirac delta function $\Theta'\left(l\right)=\delta\left(l\right)$.

For the spacetime regions $\mathcal V^\pm$ to be matched smoothly at $\Sigma$, two junction conditions must be satisfied. These are the continuity of the induced metric $h_{\alpha\beta}$ and the extrinsic curvature $K_{\alpha\beta}$,
\begin{equation}\label{defhab}
h_{\alpha\beta}=g_{ab}e^a_\alpha e^b_\beta,
\end{equation}
\begin{equation}\label{defkab}
K_{\alpha\beta}=e^a_\alpha e^b_\beta\nabla_an_b,
\end{equation}
where $\nabla_a$ denotes a covariant derivative. Defining the jump of a given quantity $X$ across the hypersurface $\Sigma$ as $\left[X\right]=X^+|_\Sigma-X^-|_\Sigma$, the junction conditions can be written as
\begin{equation}\label{junctioncond}
\left[h_{\alpha\beta}\right]=\left[K_{\alpha\beta}\right]=0,
\end{equation}
The first of these conditions comes from the fact that when one takes the derivative of the metric $g_{ab}$, written in the distribution formalism, with respect to $x^a$, terms proportional to $\delta\left(l\right)$ will arise. When one computes the Christoffel symbols, these terms must vanish because otherwise the Christoffel symbols would depend on products of the form $\Theta\left(l\right)\delta\left(l\right)$, which are not defined in the distribution formalism and thus the formalism would cease to be valid. On the other hand, the second junction condition assures that no $\delta\left(l\right)$ terms are present in the stress-energy tensor $T_{ab}$ in the field equations. 

Note that the second junction condition is not mandatory because it does not give rise to terms of the form $\Theta\left(l\right)\delta\left(l\right)$, and therefore if this condition is violated we can still perform the matching with a thin shell of matter at the hypersurface $\Sigma$. The stress-energy tensor $S_{ab}$ of the resultant thin shell can be written as
\begin{equation}\label{sabshell}
S_{\alpha\beta}=-\frac{1}{8\pi}\left(\left[K_{\alpha\beta}\right]-\left[K\right]h_{\alpha\beta}\right),
\end{equation}
where $K$ is the trace of the extrinsic curvature $K_{\alpha\beta}$. Furthermore, writing $S_\alpha^\beta=\text{diag}\left(\sigma,p_t,p_t\right)$, the surface energy density $\sigma$ and the transverse pressure $p_t$ of the thin shell can be obtained.

The induced metric $h_{\alpha\beta}$ of the hypersurface $\Sigma$ can be obtained from Eq.\eqref{defhab} and its line element takes the general form
\begin{equation}\label{indhab}
ds_\Sigma^2=-d\tau^2+R_\Sigma\left(\tau\right)^2\left(d\theta^2+\sin^2\theta d\phi^2\right),
\end{equation}
in standard spherical coordinates, where $\tau$ denotes the proper time coordinate of the hypersurface and $R_\Sigma$ is the radius at which this hypersurface stands, i.e., the matching radius between the two spacetimes $\mathcal V^\pm$. In the following sections, we will denote derivatives with respect to $\tau$ with an over dot ( $\dot{}$ ). 

\subsection{Weak, strong, and dominant energy conditions}

In the context of GR, the stress-energy tensor describing a given matter distribution, e.g. a fluid sphere like the Schwarzschild star from Sec.\ref{sub:star} or a thin shell from Sec.\ref{sub:shell}, is expected to satisfy a few properties. In particular, for physically relevant configurations, one expects the energy density to be positive and dominant over pressure. These properties are known as the energy conditions.

\subsubsection{The weak energy condition}

The WEC states that the average energy density as seen from an observer moving along a timelike vector field $v^a$ must be positive. This corresponds to a condition on the stress-energy tensor $T_{ab}$ of the form:
\begin{equation}
T_{ab}v^av^b\geq 0.
\end{equation}
For the particular case in which matter can be described by an isotropic perfect fluid, i.e., the stress energy tensor can be written in the form given in Eq.\eqref{stress}, then the WEC becomes
\begin{equation}\label{wec}
\rho\geq 0,\qquad \rho+p\geq 0.
\end{equation}
These results must be valid for any stress-energy tensor independently of its dimension, i.e., similar conditions arise for the stress-energy tensor $S_{ab}$ of a thin shell in the forms $\sigma \geq 0$ and $\sigma + p_t\geq 0$.

\subsubsection{The strong energy condition}

The SEC is more of a geometrical property instead of a matter-related one. Effectively, it states that the trace of the tidal tensor, i.e., the Ricci tensor $R_{ab}$, must be non-negative as measured by any observers moving along the same timelike vector field $v^a$. This corresponds to a condition on the stress-energy tensor $T_{ab}$ and its trace $T$ of the form
\begin{equation}
\left(T_{ab}-\frac{1}{2}Tg_{ab}\right)v^av^b\geq 0.
\end{equation}
Again, considering that the matter distribution can be well modeled by an isotropic perfect fluid, the stress-energy tensor $T_{ab}$ is given by Eq.\eqref{stress} and we obtain
\begin{equation}\label{sec}
\rho+p\geq 0,\qquad \rho + 3p\geq 0.
\end{equation}
Similarly, for the stress-energy tensor $S_{ab}$ of a thin shell these conditions become $\sigma+p_t\geq 0$ and $\sigma + 2p_t\geq 0$. Note that the SEC does not imply the WEC as the positiveness of the energy density is no longer required, and thus these two conditions must be checked independently. 

\subsubsection{The dominant energy condition}

The DEC imposes that matter moves along timelike or null world lines. In other words, for an observer moving along an arbitrary future-directed timelike vector field $v^a$, the measured matter's momentum density $-T^a_bv^b$ must also be future-directed and it must not be a spacelike vector field. In the particular case in which matter is described by an perfect fluid, i.e., the stress-energy tensor is written in the form given in Eq.\eqref{stress}, the DEC becomes
\begin{eqnarray}\label{dec}
\rho\geq 0,\qquad \rho\geq |p|.
\end{eqnarray}
For a thin shell with a stress energy tensor $S_{ab}$, these conditions become $\sigma\geq 0$ and $\sigma\geq |p_t|$. Furthermore, comparing Eqs.\eqref{dec}, \eqref{wec}, and \eqref{sec}, note that the DEC implies the WEC, but it does not imply the SEC.

\section{Static configurations}\label{sec:static}

In this section we are interested in obtaining static fluid configurations supported by thin shells. The method to compute these solutions is as follows: we start by defining the metrics that describe the interior and the exterior spacetimes, i.e., the metric $g_{ab}^\pm$. Using Eq.\eqref{defhab}, we compute the induced metrics $h_{\alpha\beta}^\pm$ as seen from the spacetimes $\mathcal V^\pm$. Since the induced metric must be continuous from the first of Eq.\eqref{junctioncond}, we equal both $h_{\alpha\beta}^\pm$ to the general form of $h_{\alpha\beta}$ described by the line element in Eq.\eqref{defhab}. This will provide constraints between the proper-time coordinates $\tau^\pm$ and the matching radii $R_\Sigma^\pm=R_\Sigma$. Afterwards, we compute the extrinsic curvature from Eq.\eqref{defkab} subjected to these constraints and we insert the results into Eq.\eqref{sabshell}. In general, the stress-energy tensor $S_{\alpha\beta}$ resultant from this calculation will depend on proper-time derivatives of $R_\Sigma$. As we are interested in static solutions, we impose $\dot R_\Sigma =\ddot R_\Sigma =0$, and we obtain the surface energy density $\sigma$ and the transverse pressure $p_t$.

Although infinitesimally thick, thin shell configurations have been shown to arise as a very good approximation to describe thick-domain walls \cite{garfinkle1, garfinkle2}. Furthermore, objects featuring thin shells (known as gravastars) have been proposed as alternative endpoints of gravitational collapse in GR \cite{mazur1}. Generally, one could think of these thin shell configurations as approximations for the layers of a fluid object where the density and pressure radial profiles change rapidly e.g. following a different equation of state.

\subsection{Model 1: Exterior thin shell}\label{sec:model1}

\begin{figure}
\centering
\includegraphics[scale=0.3]{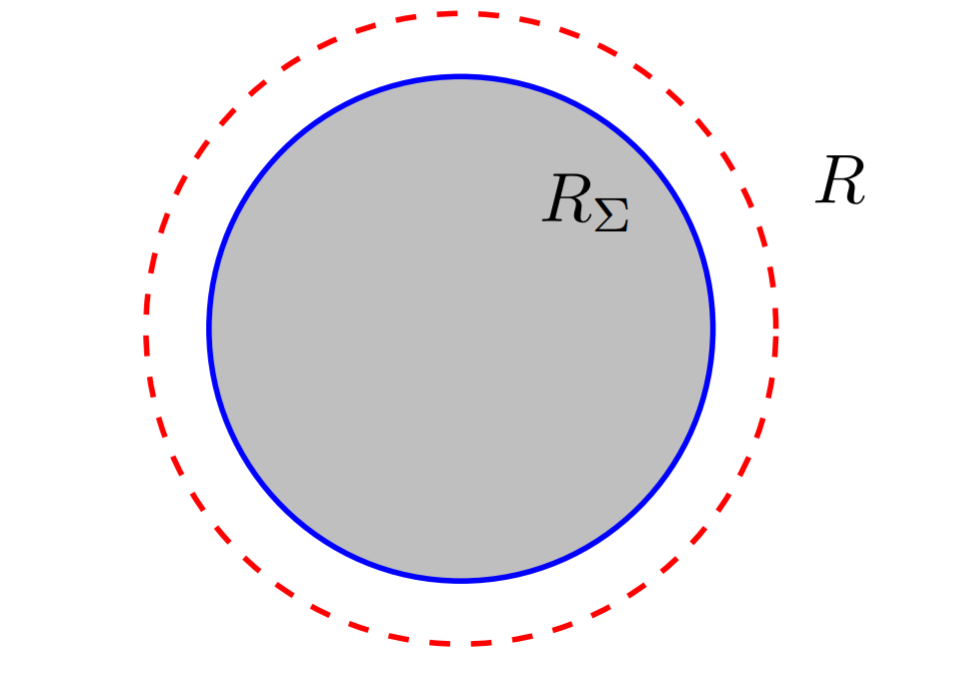}
\caption{Schematic representation of model 1: a thin shell (solid blue) separates the interior Schwarzschild spacetime (light gray) from the exterior Schwarzschild spacetime (white) at a radius $R_\Sigma$ smaller than the radius of the star $R$ (dashed red).}
\label{fig:extmatch}
\end{figure}

In this section, we shall take a Schwarzschild star described by the interior and exterior metrics provided in Eqs.\eqref{metricint} and \eqref{metricext}, respectively, with a radius $R$ greater than the Buchdahl limit, and perform a matching between the two at a junction radius $R_\Sigma<R$. As the extrinsic curvature $K_{ab}$ is no longer continuous across $\Sigma$, a thin shell will arise at the junction radius, see Fig.\ref{fig:extmatch}. 

Since we have taken the radius of the star to be greater than the Buchdahl limit, $R>R_b$, this implies from the results of Sec.\ref{sub:star} that the pressure $p$ of the interior fluid is finite and monotonically decreasing throughout the whole interior solution, and also that no singularities are present, independently of the junction radius. 

\begin{figure*}
\centering
\includegraphics[scale=0.6]{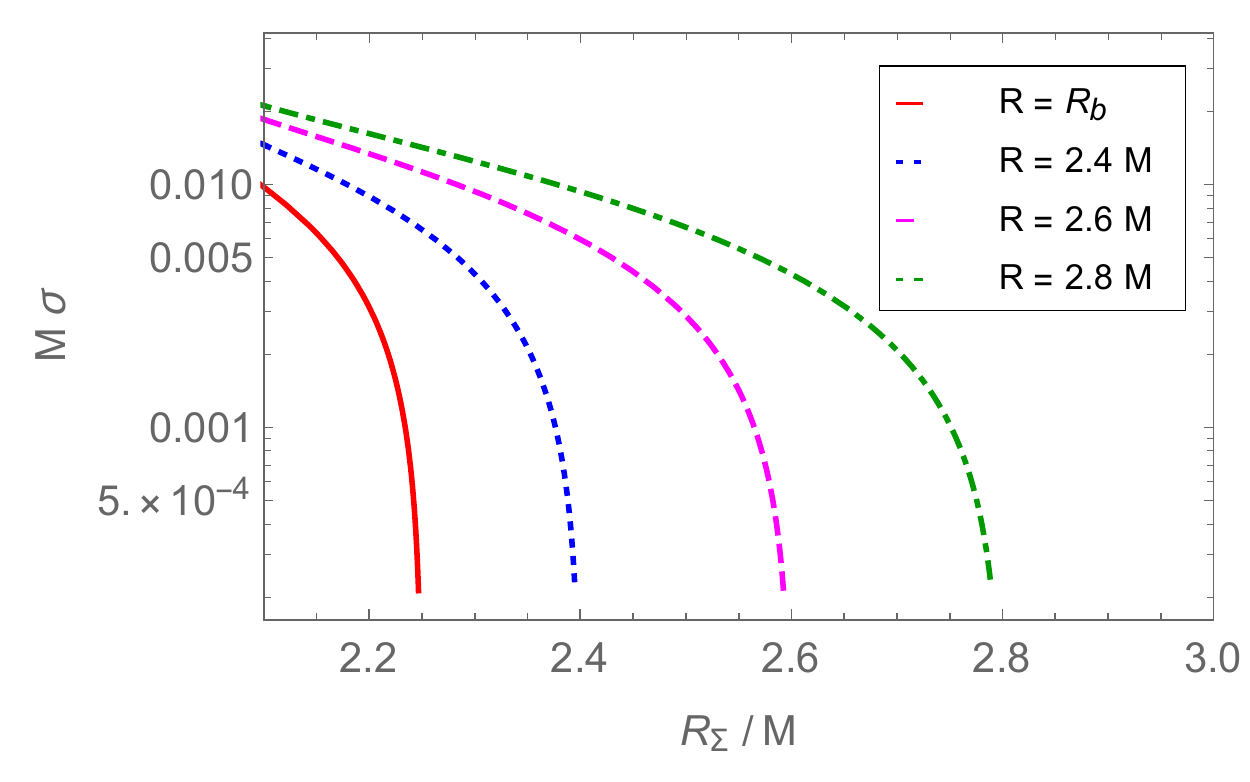}
\ \ \ \ \ \ \ \ \ \ 
\includegraphics[scale=0.6]{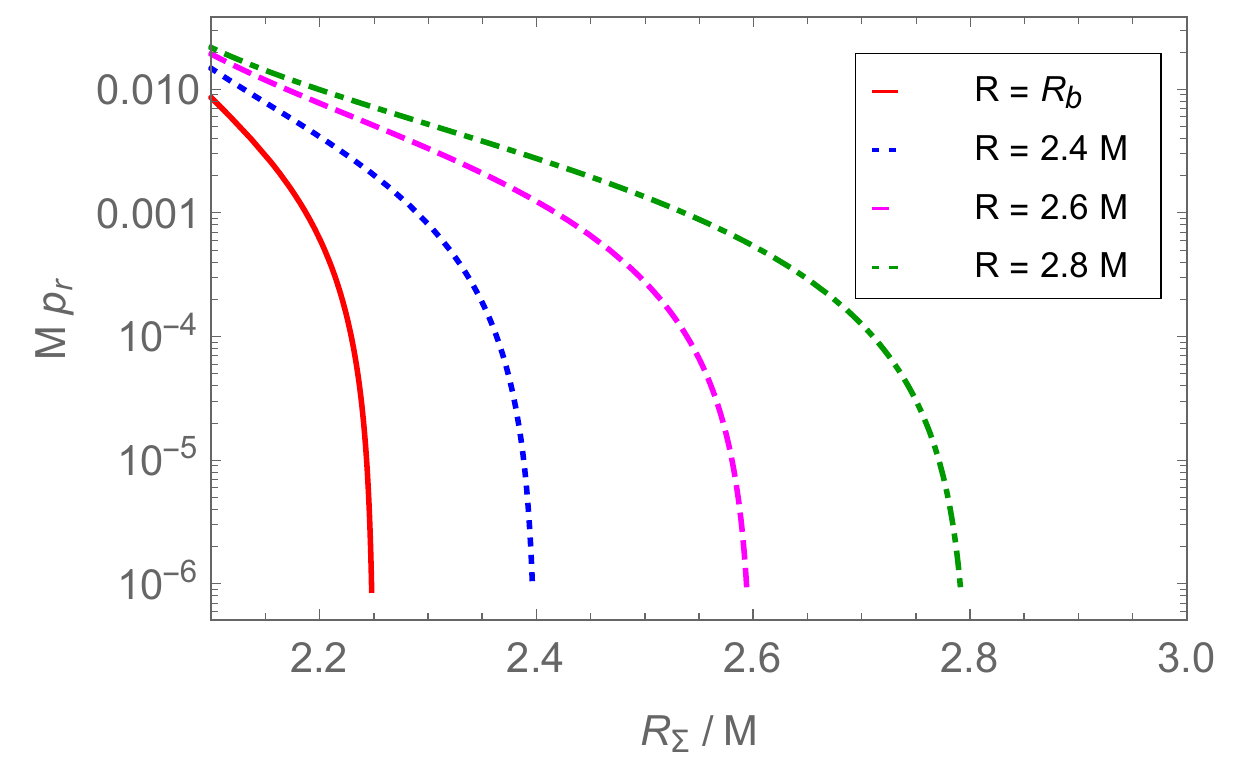}
\caption{Normalized density $M\sigma$ and normalized surface pressure $M p_t$ for the model depicted in Fig.\ref{fig:extmatch} as a function of the junction radius $R_\Sigma$ for different star radii $R$. As expected, both $\sigma$ and $p_t$ vanish when $R_\Sigma=R$, thus recovering a smooth matching between the two spacetime regions.}
\label{fig:matter1}
\end{figure*}

As can be seen from Eq.\eqref{sabshell}, as we vary the junction radius, and consequently the jump of the extrinsic curvature, the stress-energy tensor $S_{ab}$ of the thin shell will change, and both the surface energy density $\sigma$ and transverse pressure $p_t$ of the thin shell will depend on the junction radius. In Fig.\ref{fig:matter1} we plot the normalized density $M\sigma$ and the normalized surface pressure $M p_t$ as a function of the junction radius $R_\Sigma$ for different values of the star radius $R$. We avoid writing the explicit dependencies of $\sigma$ and $p_t$ in the junction radius $R_\Sigma$ due to their size.

\begin{figure*} 
    \begin{center}
    \centering
    \includegraphics[scale=0.5]{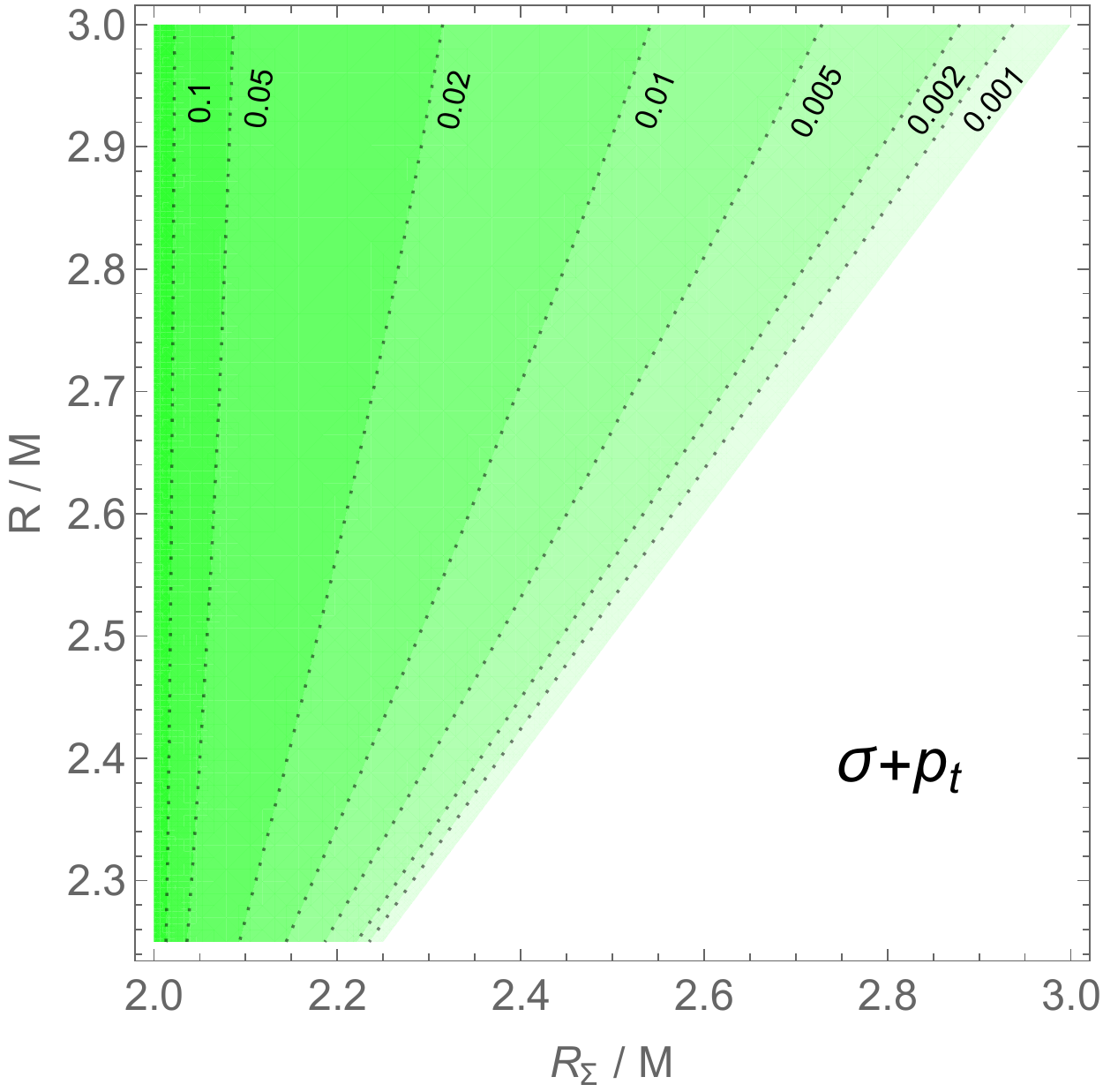}
    \ \ \ \ \ \ \ \ \ \ 
    \includegraphics[scale=0.5]{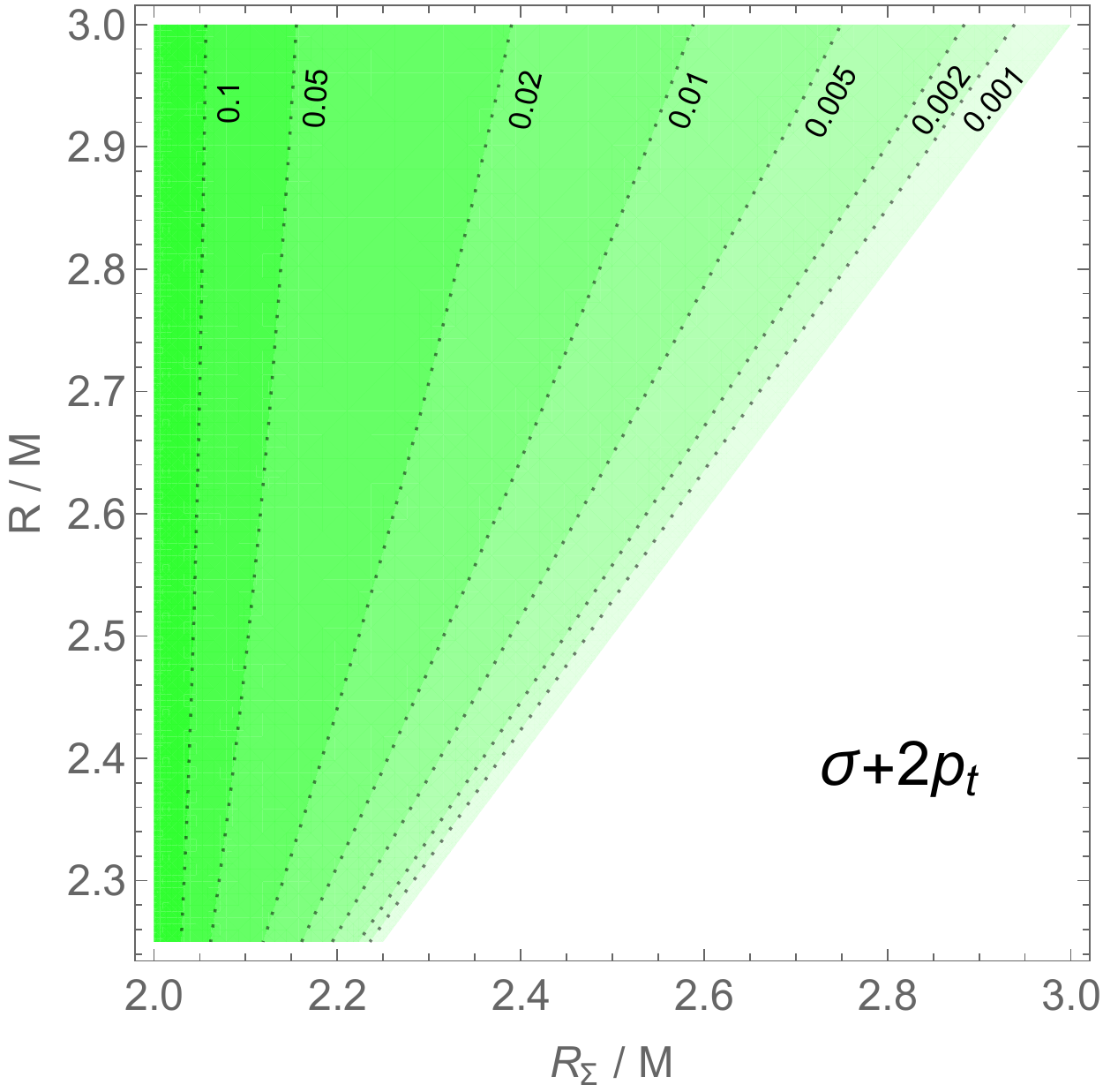}
    \caption{Parameter space of $R$ and $R_\Sigma$ for the model depicted in Fig.\ref{fig:extmatch}. In the left panel we plot $M\left(\sigma+p_r\right)$, positive where the WEC is satisfied, whereas in the right panel we plot $M\left(\sigma+2p_r\right)$, positive where the SEC is satisfied. Both the WEC and the SEC are satisfied regardless of the junction radius $R_\Sigma$ as long as $R>R_b$.}
    \label{fig:ecext}
    \end{center}
    \end{figure*}

Regardless of the value of the junction radius, we verify that $\sigma>0$, as expected since we are collapsing the outer layers of the star in the thin shell, and also $p_t>0$. Consequently, both the WEC and the SEC, given in Eqs.\eqref{wec} and \eqref{sec} respectively, are automatically satisfied, see Fig.\ref{fig:ecext}. 

These results imply that we can perform the matching between the two spacetimes arbitrarily close to the Schwarzschild radius $R=2M$ and obtain a model for an incompressible and isotropic relativistic fluid sphere that does not develop singularities, thus being an exception to the Buchdahl's limit, and that still satisfies both the WEC and the SEC.

\subsection{Model 2: Interior thin shell}\label{sec:model2}

\begin{figure}
\centering
\includegraphics[scale=0.3]{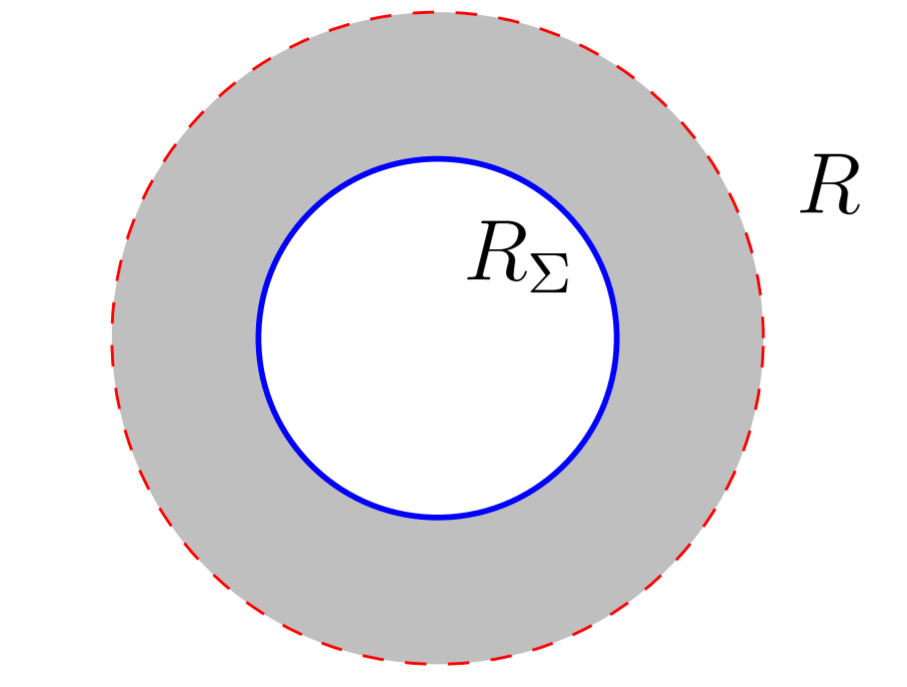}
\caption{Schematic representation of model 2: a thin shell (solid blue) separates the interior Minkowski spacetime (white) from the interior Schwarzschild spacetime (light gray) at a radius $R_\Sigma$ smaller than the radius of the star $R$ (dashed red).}
\label{fig:intmatch}
\end{figure}

Let us now consider an alternative approach to the problem. Again, take a Schwarzschild star described by the interior and exterior metrics given in Eqs.\eqref{metricint} and \eqref{metricext}, respectively, but now we let the radius of the star be smaller than the Buchdahl limit, i.e., $R<R_b$. According to the results from Sec.\ref{sub:star}, this implies that we will have a singularity in the interior fluid region at some radius $R_d$. To overcome this problem, let us create a "vacuum bubble" in the central region of the star, described by the Minkowski metric
\begin{equation}
ds^2=-dt^2+dr^2+r^2\left(d\theta^2+\sin^2\theta d\phi^2\right),
\end{equation}
and perform a matching between the Minkowski spacetime and the interior Schwarzschild spacetime at a given matching radius $R_\Sigma$, see Fig.\ref{fig:intmatch}.

If we choose an adequate value for $R_\Sigma$, i.e., greater than the radius for which the singularity occurs, $R_\Sigma>R_d$, then this feature is effectively removed from the model. From Eqs.\eqref{pressure} and \eqref{metricint}, we verify that 
\begin{equation}\label{divrad}
R_d=3R\sqrt{1-\frac{4R}{9M}}.
\end{equation}
As a consequence, the pressure $p$ is again finite and monotonically decreasing in the outwards radial direction inside the fluid region.

\begin{figure*}[]
    \centering
    \includegraphics[scale=0.6]{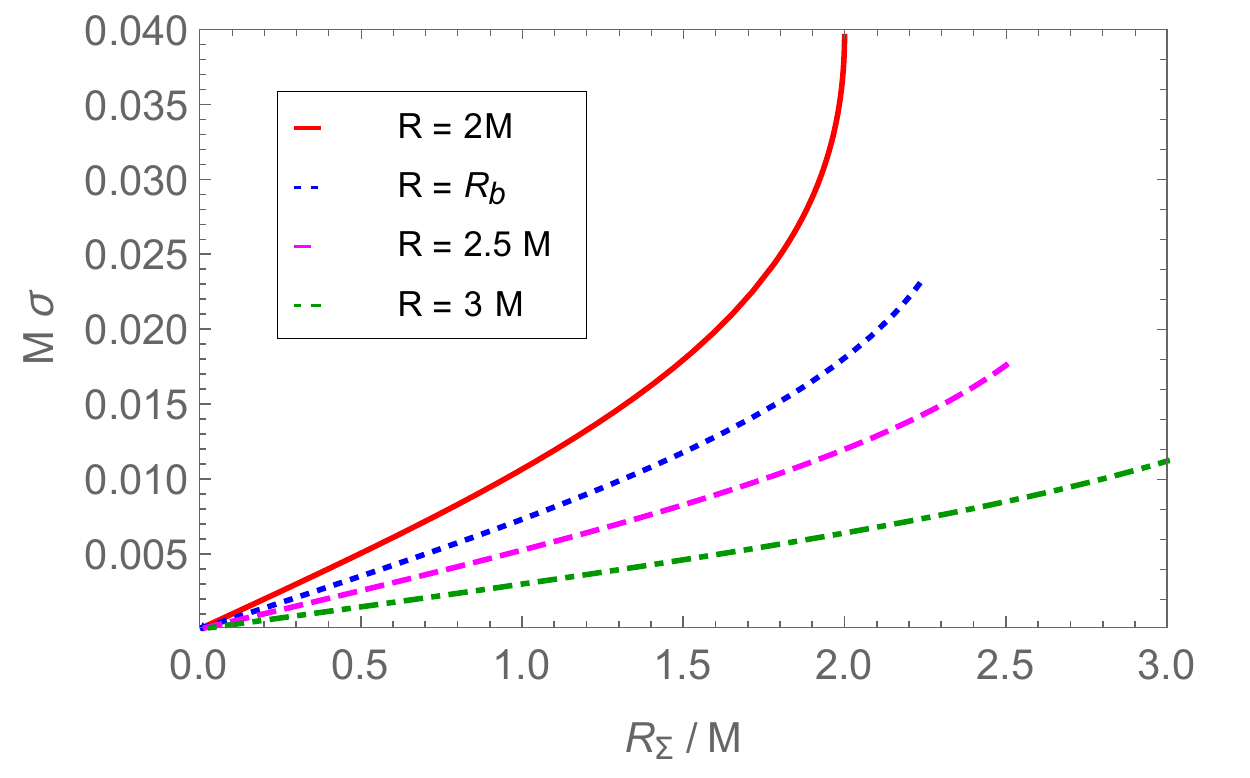}
    \ \ \ \ \ \ \ \ \ \ 
    \includegraphics[scale=0.6]{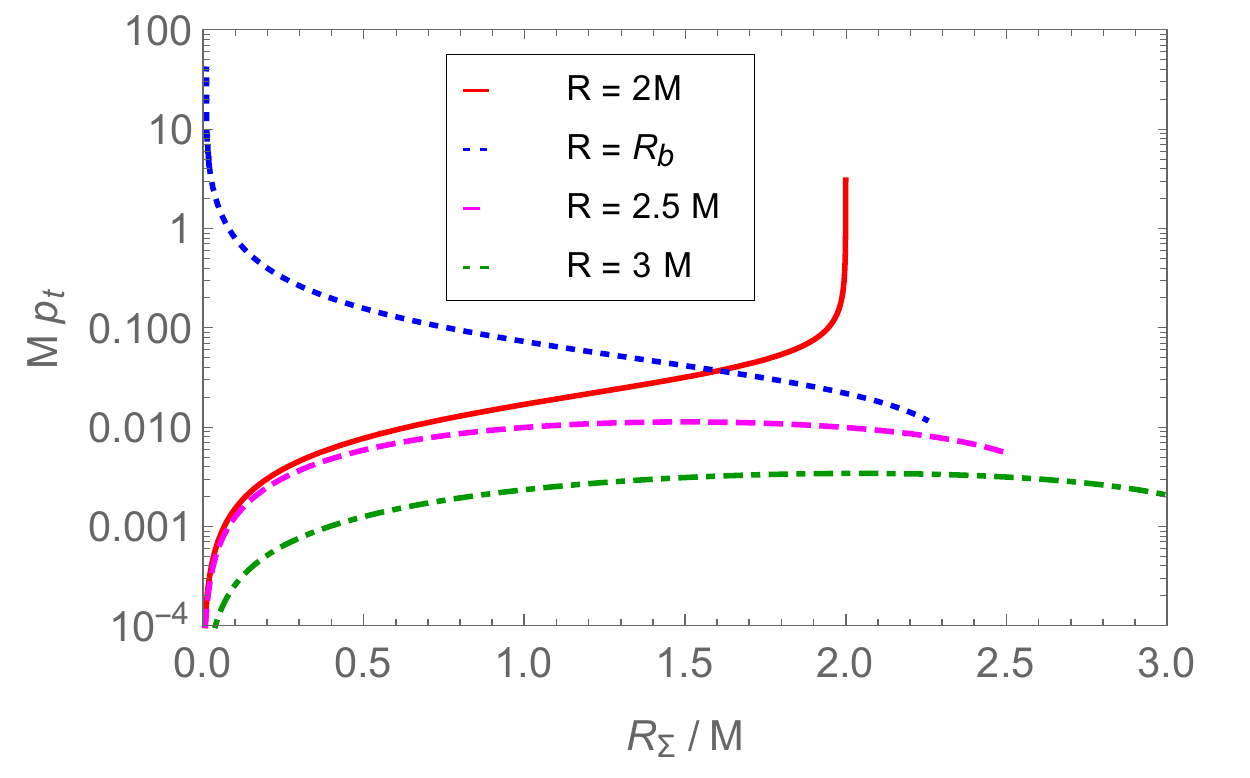}
    \caption{Normalized density $M\sigma$ and absolute value of the normalized surface pressure $M p_t$ for the model depicted in Fig.\ref{fig:intmatch} as a function of the junction radius $R_\Sigma$ for different star radii $R$. As expected, both $\sigma$ and $p_t$ vanish when $R_\Sigma=0$ for $R>R_b$. However, when $R=R_b$, we verify that $p_t$ diverges at the origin. Furthermore, for $R=2M$, we have $p_t<0$ (the solid red line plots the absolute value of this quantity) and it diverges at $R_\Sigma=2M$.}
    \label{fig:matter2}
\end{figure*}

Similarly to the previous model, due to Eq.\eqref{sabshell} the surface density $\sigma$ and the transverse pressure $p_t$ of the thin shell will depend on the junction radius $R_\Sigma$. These dependencies are again very lengthy so we chose not to write them explicitly. In Fig.\ref{fig:matter2} we plot the normalized energy density $M\sigma$ and the normalized surface pressure $Mp_t$ as a function of the junction radius $R_\Sigma$ for different values of the star radius $R$.

\begin{figure*}
    \begin{center}
    \centering
    \includegraphics[scale=0.5]{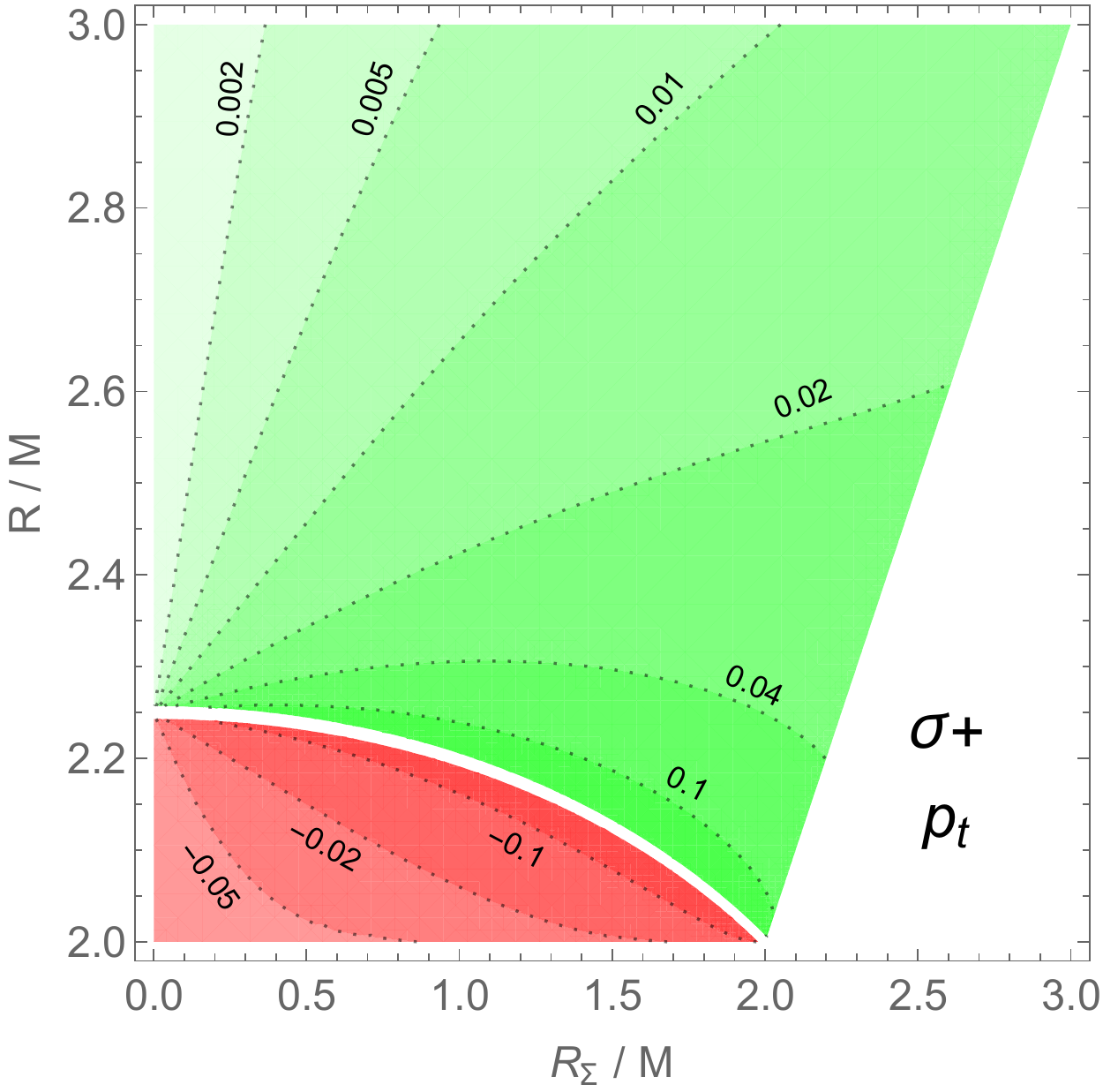}
    \ \ \ \ \ \ \ \ \ \ 
    \includegraphics[scale=0.5]{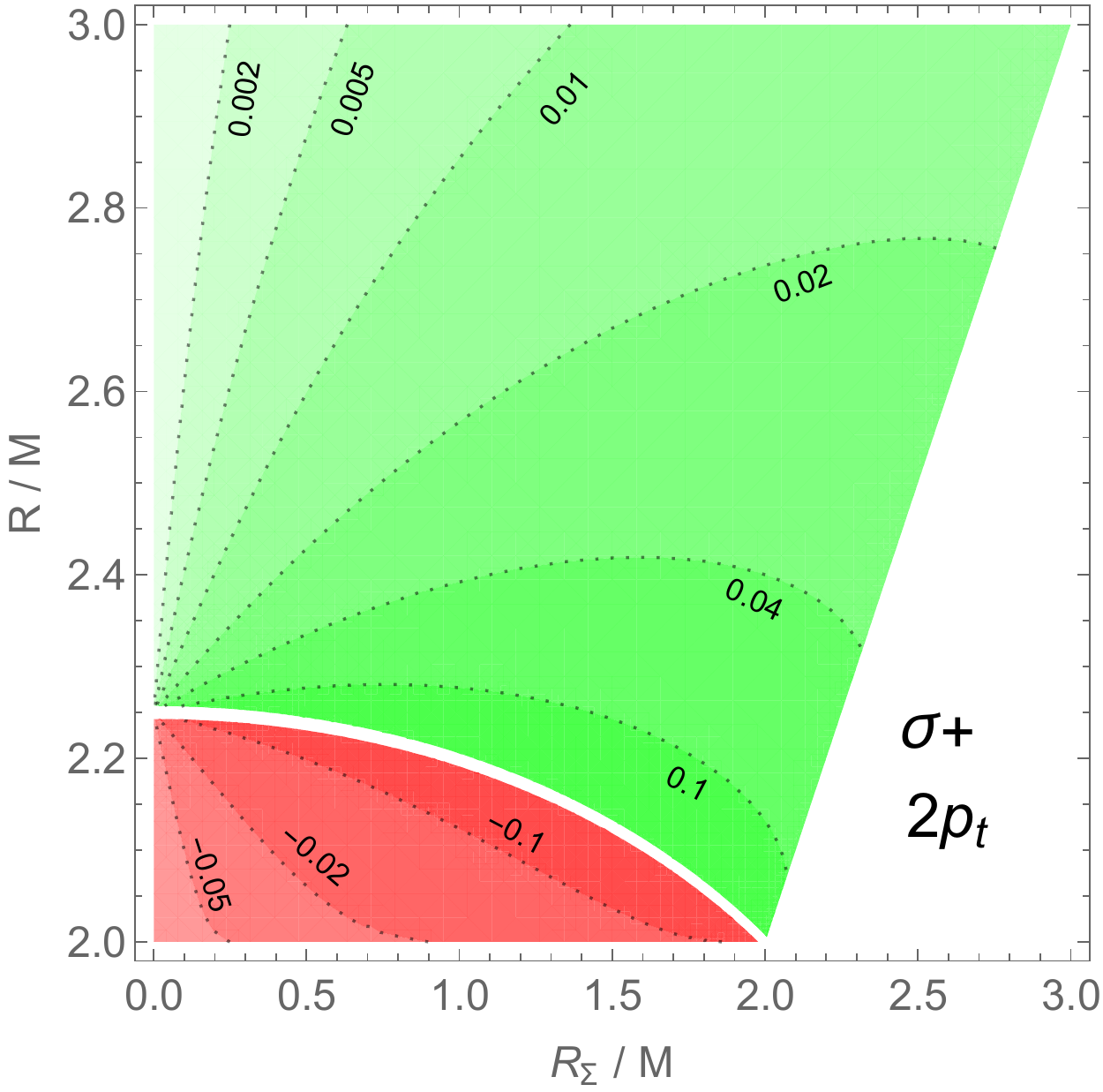}
    \caption{Parameter space of $R$ and $R_\Sigma$ for the model depicted in Fig.\ref{fig:intmatch}. In the left panel we plot $M\left(\sigma+p_t\right)$, positive where the WEC is satisfied, whereas in the right panel we plot $M\left(\sigma+2p_t\right)$, positive where the SEC is satisfied. Both the WEC and the SEC are satisfied as long as $R_\Sigma>R_d$.}
    \label{fig:ecint}
    \end{center}
    \end{figure*}

Independently of the radius $R_\Sigma$, we verify that $\sigma>0$, which is again an expected result as this would correspond to collapsing the inner layers of the star outward into a thin shell. However, $p_t$ can be negative for some choices of $R$ and $R_\Sigma$, resulting in a consequent violation of both the WEC and the SEC for some regions of the parameter space, see Fig.\ref{fig:ecint}.

These results show a smooth separation between the region where the energy conditions are satisfied from the region where they are violated. Curiously, the line that separates the two regions corresponds exactly to $R_\Sigma=R_d$ in Eq.\eqref{divrad}. This implies that as long as we perform the matching at a radius $R_\Sigma$ large enough to remove the singularity from the interior solution, the stress-energy tensor of the thin shell will automatically satisfy both the WEC and the SEC. Therefore, we obtain another model for an incompressible and isotropic relativistic fluid configuration for which the radius can be arbitrarily close to the Schwarzschild radius $R=2M$ without developing singularities.

\section{Stability analysis}\label{sec:stability}

\subsection{Dynamical thin shell framework}

In the previous sections, as we were looking for static thin shell configurations, we have imposed the equilibrium conditions $\dot R_\Sigma = \ddot R_\Sigma=0$ before computing the surface energy density $\sigma$ and the surface pressure $p_t$ of the shell. Here, we are interested in studying the stability of these systems, and thus we shall drop these assumptions and consider the framework of dynamical thin shells\cite{crawford,garcia1}. 

\subsubsection{Equation of motion for the thin shell}

In comparison with the previous approach, the main difference arising from dropping the staticity assumption is that a dependency in the proper-time derivatives of the matching radius, $\dot R_\Sigma$ and $\ddot R_\Sigma$ appears in the induced metric $h_{\alpha\beta}$, the extrinsic curvature $K_{\alpha\beta}$, and consequently in the stress-energy tensor of the thin shell $S_{\alpha\beta}$. In particular, the equation for the surface energy density $\sigma$, i.e., the $(0,0)$ component of the stress-energy tensor of the thin shell given in Eq.\eqref{sabshell}, becomes a function of $\dot R_\Sigma$. This allows us to deduce an equation of motion for the thin shell in the form
\begin{eqnarray}
\dot R_\Sigma +V\left(R_\Sigma\right)=0,
\end{eqnarray}
where $V\left(R_\Sigma\right)$ is the thin shell potential, written in terms of the metrics $g_{ab}^\pm$ and the mass of the thin shell $m_s=4\pi \sigma R_\Sigma^2$ as
\begin{equation}\label{shellpot}
V\left(R_\Sigma\right)=F\left(R_\Sigma\right)-\left(\frac{m_s}{2R_\Sigma}\right)^2-\left(\frac{G\left(R_\Sigma\right)R_\Sigma}{m_s}\right)^2,
\end{equation}
where the functions $F\left(R_\Sigma\right)$ and $G\left(R_\Sigma\right)$ are respectively the average and the symmetric of the half of the jump of the inverse metric components $g_{rr}^\pm$ across the hypersurface $\Sigma$:
\begin{equation}
F\left(R_\Sigma\right)=\frac{1}{2}\left(\frac{1}{g_{rr}^-}+\frac{1}{g_{rr}^+}\right),
\end{equation}
\begin{equation}
G\left(R_\Sigma\right)=\frac{1}{2}\left(\frac{1}{g_{rr}^-}-\frac{1}{g_{rr}^+}\right).
\end{equation}

The stability problem of a thin shell is thus similar to that of the stability of a particle moving in a one-dimensional potential. We expand the potential in a Taylor series around the equilibrium radius for the static solutions computed previously, $R_\Sigma=R_0$, from which we can immediately verify that $V\left(R_0\right)=0$. Furthermore, as $R_\Sigma=R_0$ is an equilibrium state, we know that $V'\left(R_0\right)=0$, where a prime denotes a derivative with respect to the junction radius $R_\Sigma$, which can also be verified taking the derivative of Eq.\eqref{shellpot}. Assuming small radial perturbations, i.e., $|R_\Sigma-R_0|\ll 1$, we are left with a potential given by
\begin{equation}
V\left(R_\Sigma\right)=\frac{1}{2}V''\left(R_0\right)\left(R_\Sigma-R_0\right)^2+\mathcal O\left(3\right),
\end{equation}
to the leading order in $R_\Sigma-R_0$. The stability regimes for a thin shell configuration are now evident: the system will be stable whenever $V''\left(R_0\right)>0$ and unstable otherwise. The term $V''\left(R_\Sigma\right)$, being a second derivative of the potential given in Eq.\eqref{shellpot}, will depend on radial derivatives of the surface energy density $\sigma$, which must be computed using the stress-energy tensor conservation equation.

\subsubsection{Conservation of the stress-energy tensor}

The conservation equation for the stress-energy tensor of the thin shell is given by the expression
\begin{equation}\label{conseqsab}
e^a_\alpha\nabla_aS^\alpha_\beta=\left[T_{ab}e^a_\beta n^b\right].
\end{equation}
In the static cases considered before, the surface energy density $\sigma$ and the surface pressure $p_t$ of the thin shell did not vary, and thus the conservation equation was automatically satisfied. In the dynamical framework, the conservation equation is an extra constraint one must take into consideration. Equation \eqref{conseqsab} can be written in the form
\begin{equation}\label{consigma}
\sigma'=-\frac{2}{R_\Sigma}\left(\sigma+p_t\right)+\Xi,
\end{equation} 
where $\Xi$ corresponds to the discontinuity in the momentum flux across the shell and can be written in terms of the metric components and their derivatives evaluated at the static solution $r=R_0$ in the general form
\begin{equation}\label{deflux}
\Xi=\frac{1}{8\pi R_\Sigma}\left[\left(\frac{g_{rr}'}{g_{rr}}-\frac{g_{tt}'}{g_{tt}}\right)\sqrt{\frac{1}{g_{rr}}+\dot R_\Sigma}\right].
\end{equation}
The conservation equation is particularly useful in this framework to rewrite the radial derivatives of the surface energy density arising in $V''\left(R_\Sigma\right)$, i.e., $\sigma'\left(R_\Sigma\right)$ and $\sigma''\left(R_\Sigma\right)$ in terms of $\sigma$ and $p$. Finally, one uses the definition of surface pressure $p_t$, i.e., the $\left(1,1\right)$ component of Eq.\eqref{sabshell} to obtain an expression for $V''\left(R_\Sigma\right)$ written solely in terms of the junction radius $R_\Sigma$ and the star radius $R$. 

\subsection{Analysis of the results}

\subsubsection{Stability regimes}

\begin{figure*}
    \begin{center}
    \centering
    \includegraphics[scale=0.55]{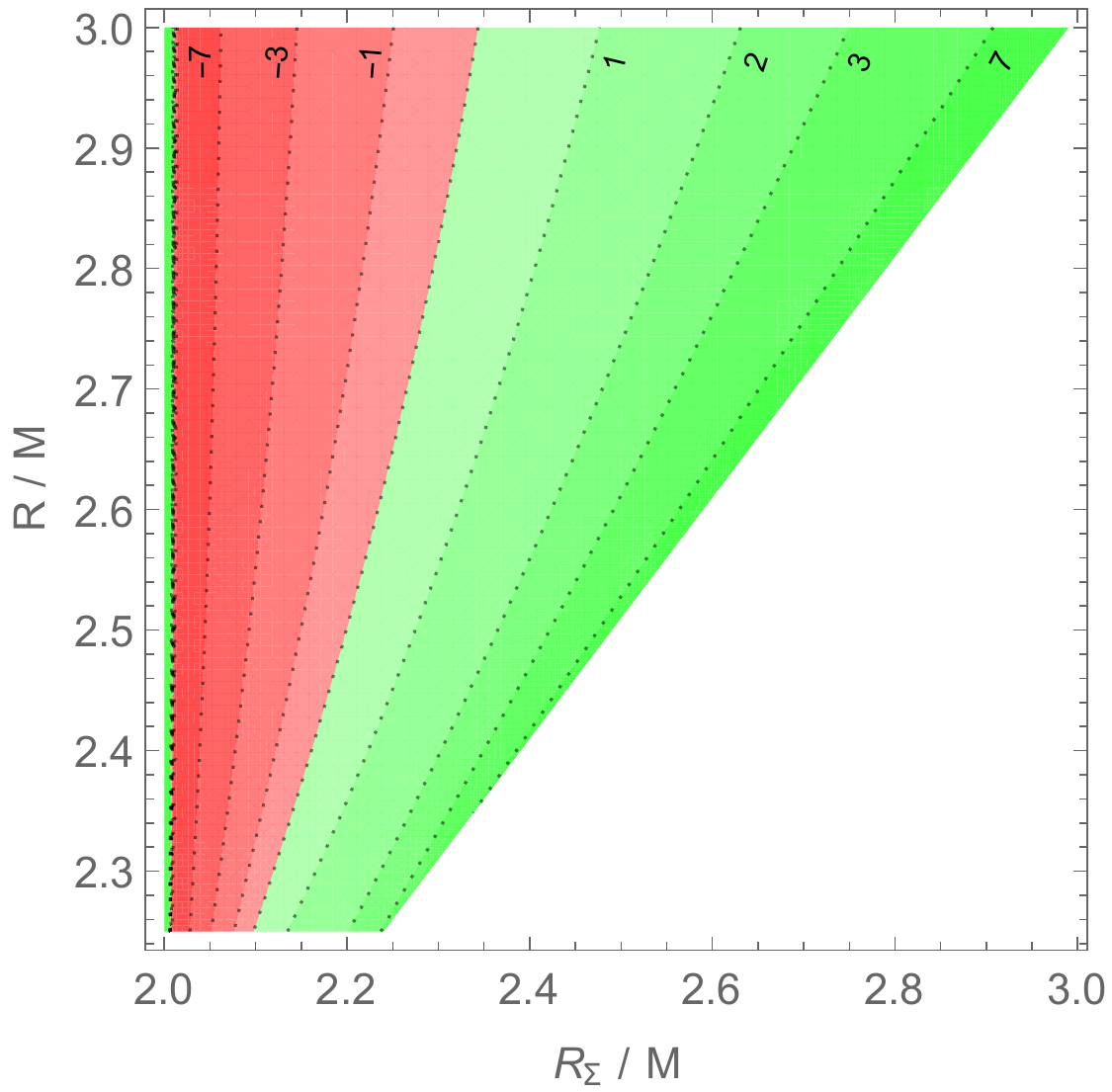}
    \ \ \ \ \ \ \ \ \ \ 
    \includegraphics[scale=0.55]{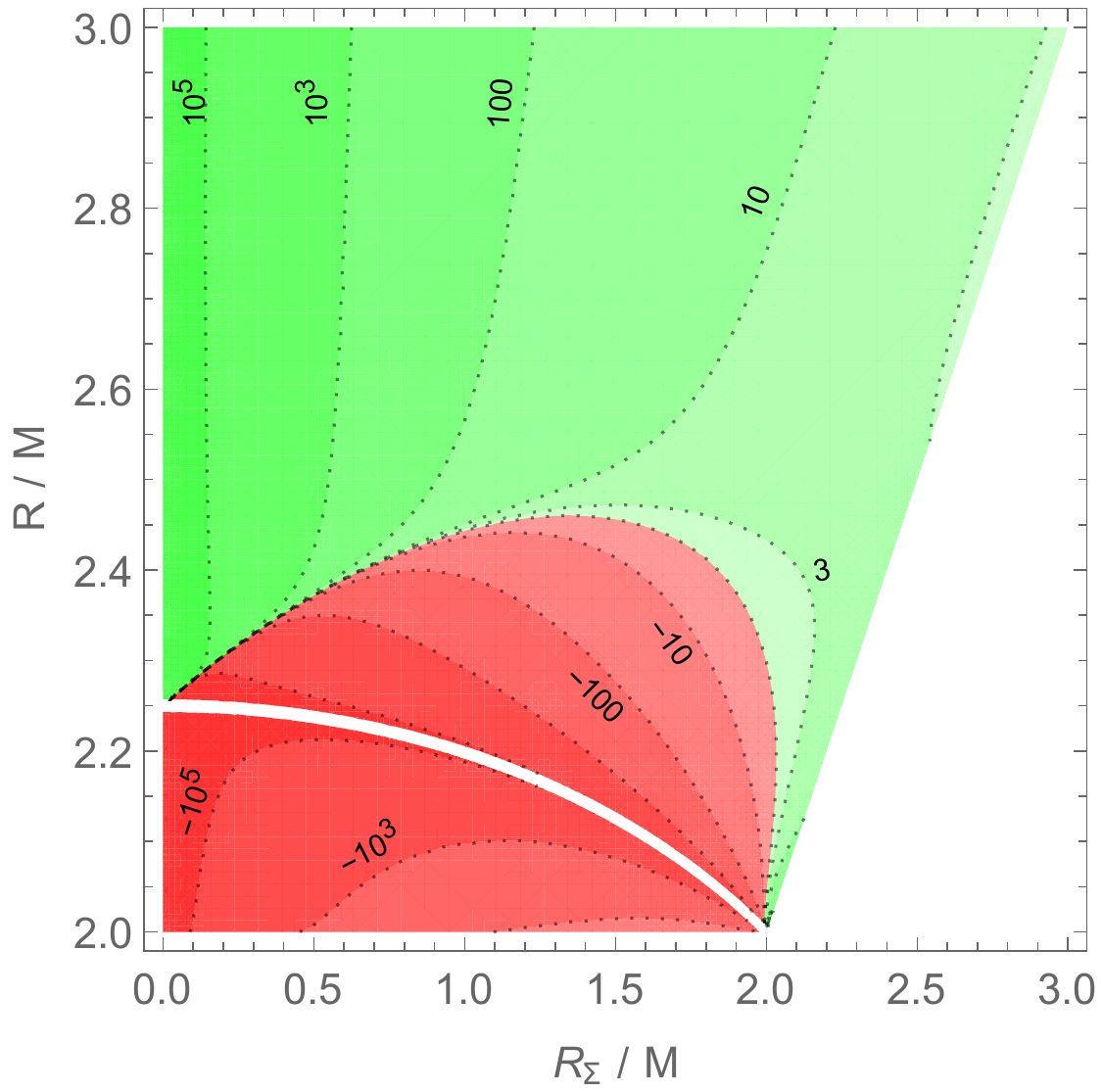}
    \caption{Parameter space of $R$ and $R_\Sigma$ for the model depicted in Fig.\ref{fig:extmatch} (left panel) and the model depicted in Fig.\ref{fig:intmatch} (right panel). We plot $V''\left(R_\Sigma\right)$, positive whenever the solution is stable and negative otherwise. We verify that stable solutions exist with an object radius smaller than $3M$ for both models and in the same regions where the WEC and the SEC are satisfied.}
    \label{fig:potential}
    \end{center}
    \end{figure*}

The stability regimes for the models proposed can be obtained by computing the second derivative of the potential provided in Eq.\eqref{shellpot} and verifying in which regions of the parameter space of $R_\Sigma$ and $R$ it is positive. As the forms of this potential and its derivatives are extremely long, we choose not to write their explicit forms. Instead, in Fig.\ref{fig:potential} we plot $V''\left(R_\Sigma\right)$ in the parameter space considered.

In both cases, one verifies that there exist combinations of the parameters $R_\Sigma$ and $R$ for which the solutions obtained are stable. However, the stability regions are clearly different from the regions where the WEC and the SEC are satisfied (see Figs.\ref{fig:ecext} and \ref{fig:ecint}). For the first model we verify that given a value for the star radius $R$, the solutions will be stable whenever the matching radius $R_\Sigma$ is greater than some critical value, which is $R_\Sigma\sim 2.1M$ for $R\sim R_b$. Interestingly, there is an extra stability region valid for any value of the star radius $R>R_b$, corresponding to a matching radius near the Schwarzschild radius, i.e. $R_\Sigma\sim 2M$. For the second model, we verify that the solutions will be stable mostly everywhere except for a limited region of the parameter space close to the divergence line $R_\Sigma=R_d$.

These results imply that for the first model we can obtain solutions for stable relativistic spheres supported by thin shells with a compactness arbitrarily close to that of a black-hole without violating both the WEC and the SEC for any initial star radius $R$. On the other hand, for the second model, there is a small region of the parameter space near $R_\Sigma\gtrsim 2M$ for which not only the solution obtained is stable but also the radius of the resultant object can be arbitrarily close to the Schwarzschild radius. However, these solutions become unstable if one considers a matching radius $R_\Sigma<2M$. 

Furthermore, for both of the models proposed, these results imply the existence of a wide variety of stable configurations with radii smaller than the light-ring radius $3M$. These correspond to solutions presenting shadows and thus they model objects that are indistinguishable from black-holes as seen from exterior observers, with the advantage of not having neither singularities or event horizons. Consequently, they constitute viable models for ECOs (in particular, black-hole mimickers) and dark matter.

\subsubsection{Validity of the DEC}

The analysis of dynamical thin shells in black-hole spacetimes has been done and provided interesting results \cite{frauendiener1,brady1}. In particular, it has been shown that all stable thin shell configurations in these backgrounds satisfy the DEC, even when its validity is not imposed \textit{a priori}. Thus, one could expect that our static models would be unstable against radial perturbations whenever the DEC is violated. 

\begin{figure*} 
    \begin{center}
    \centering
    \includegraphics[scale=0.5]{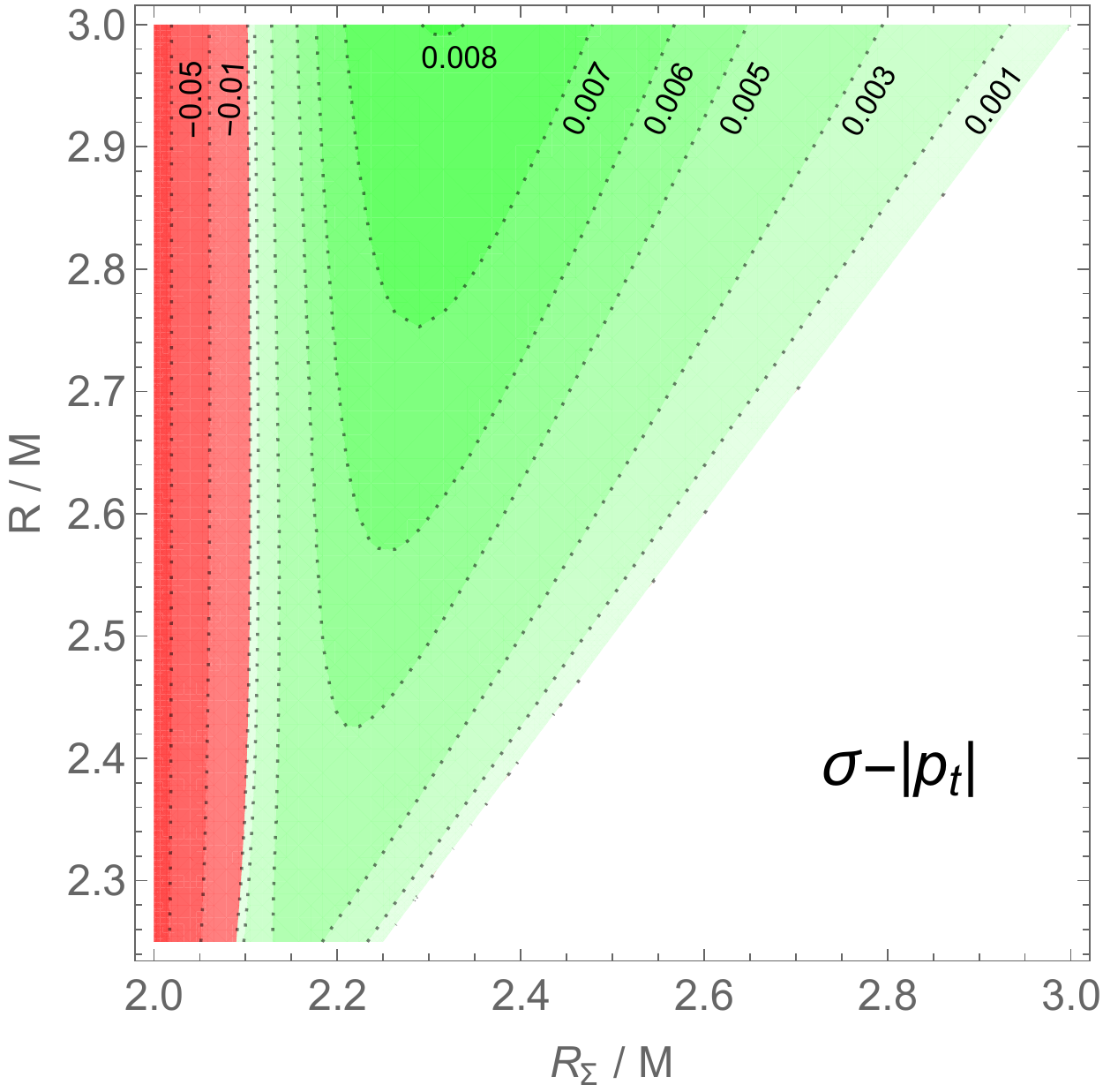}
    \ \ \ \ \ \ \ \ \ \ 
    \includegraphics[scale=0.5]{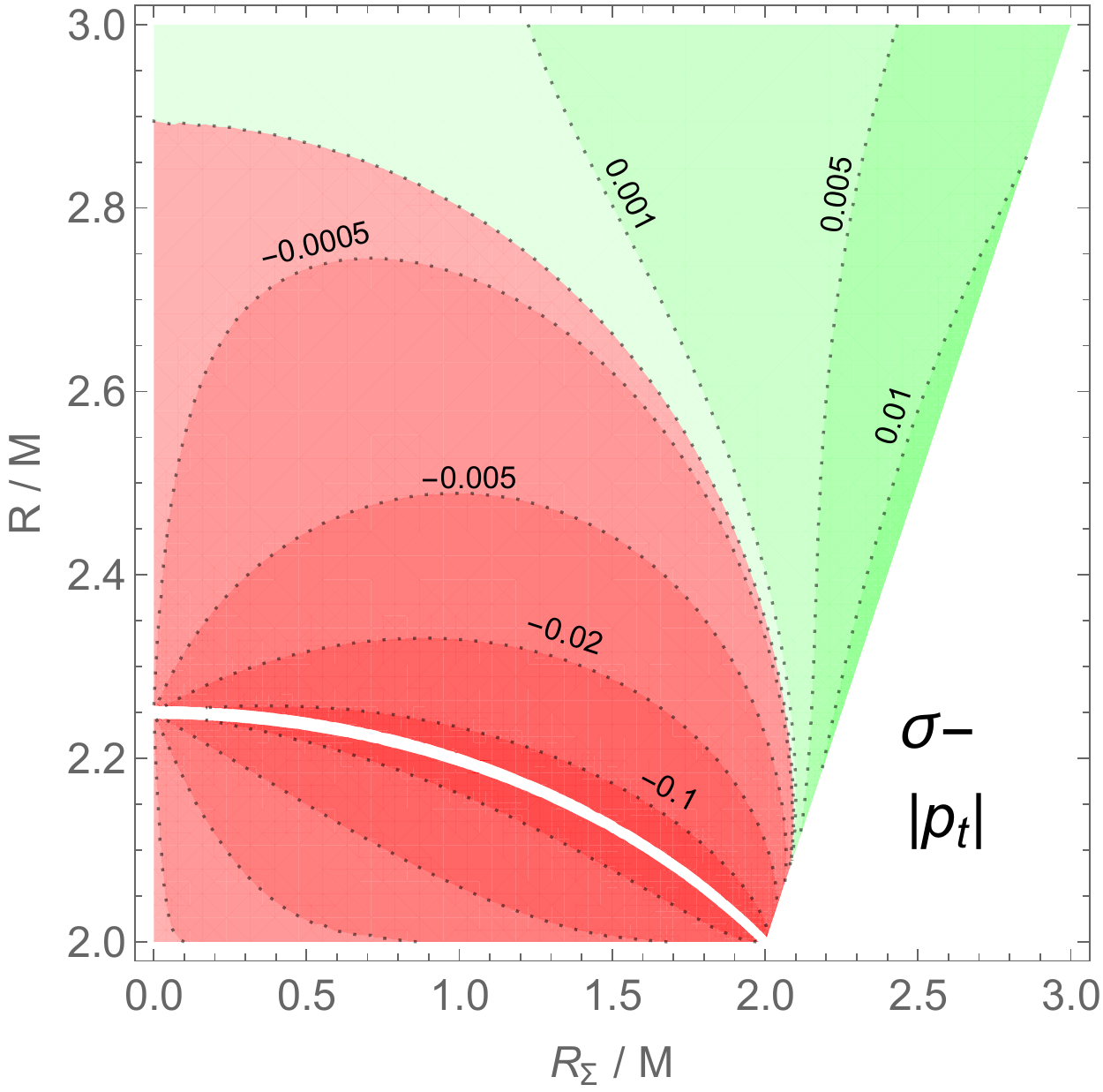}
    \caption{Parameter space of $R$ and $R_\Sigma$ for the model depicted in Fig.\ref{fig:extmatch} (left panel) and the model depicted in Fig.\ref{fig:intmatch} (right panel). We plot $M\left(\sigma-|p_t|\right)$, positive where the DEC is satisfied. For both models, the DEC is violated in a region where the WEC and the SEC are satisfied. However, there are still regions where the DEC is satisfied and the resultant object still presents a radius smaller than $R_b$.}
    \label{fig:dec}
    \end{center}
    \end{figure*}

In Fig.\ref{fig:dec}, we plot the validity regions of the DEC, i.e., we plot $M\left(\sigma-|p_t|\right)$, for both of the models proposed in Secs.\ref{sec:model1} and \ref{sec:model2}. For both models, we verify that there are regions of the parameter space for which the DEC is violated, even if the WEC and the SEC are satisfied. A comparison between Fig.\ref{fig:potential} and Fig.\ref{fig:dec} reveals something unexpected. In both of the models, there are regions of the parameter space of $R_\Sigma$ and $R$ for which the DEC is violated but the solutions remains stable nevertheless.  

\subsubsection{Implications for GW physics}

The fact that the models proposed do not present event horizons has important implications to the physics of gravitational waves. In particular, there will be effects on the GW signal if the object resultant from a coalescence of two compact objects is one of the models proposed, or if one of this objects is present in a coalescing binary.

The ringdown phase of the GW signal is dominated by the frequencies of the proper oscillation modes of the resultant object, known as the quasinormal modes, or QNMs. There frequencies are well known for black holes. If the resultant object from the coalescence is not a black-hole but some ECO instead, these frequencies will differ and will be potentially detectable via GW spectroscopy. Furthermore, the absence of an event horizon will allow for a GW to be reflected inside the object and consequently re-emitted, giving rise to periodic structures in the ringdown called echoes \cite{vitor3,vitor2}. A perturbative analysis of these models allows one to extract their proper oscillation frequencies as well as the echoes structure.

On the other hand, it is well-known that the tidal Love numbers of black-holes vanish identically \cite{binnington1} . However, the same is not true for horizonless objects such as neutron stars \cite{hinderer1} or horizonless ECOs with and without thin shells \cite{franzin1}. These tidal effects, more precisely tidal deformability and tidal heating, appear as a fifth-order post-Newtonian correction on the phase of the GW waveform, thus being potentially detectable.

\section{Conclusions}\label{sec:concl}

In this work, we have used the junction conditions and the thin shell formalism in GR to construct two models for relativistic fluid spheres supported by thin shell configurations. These models can present a compactness arbitrarily close to that of a black hole without developing singularities, thus being exceptions to the Buchdahl theorem. Furthermore, we have analyzed the validity of both the WEC and the SEC and verified that they are satisfied.

In the first model, we have shown that if we perform the matching between the Schwarzschild interior and exterior solutions for a junction radius smaller than the radius of the star, it is possible to contract the object below the Buchdahl limit while keeping the energy density constant in the interior solution, and the resultant thin shell at the junction radius will satisfy the WEC and the SEC. For the second model, we have shown that there are regions in the parameters space where the WEC and the SEC are violated, but this corresponds to the same parameter region for which singularities exist. If the junction radius is chosen to be greater than the radius at which this problem arises, the singularities are removed and the WEC and the SEC are automatically satisfied.

The stability of these objects was also analyzed and we have verified that stable solutions with a radius smaller than the radius of the light-ring, i.e., $R<3M$, exist. In particular, for the first model we have shown that stable solutions with a compactness arbitrarily close to that of a black hole exist for any initial star radius $R$, whereas for the second model there exist combinations of the parameters $R_\Sigma$ and $R$ for which the solutions are stable and present radii arbitrarily close to the Schwarzschild radius. We also show that the validity of the DEC is not a necessary condition for the stability of the second model.

The models proposed in this work correspond to objects presenting a shadow and consequently indistinguishable from black-holes as seen from an exterior observer. Although these solutions should be regarded as toy models, they can thus provide important insights for relevant candidates for dark matter and ECOs. Furthermore, it is expected that the continuous increase in the sensitivity of the gravitational wave observatories will allow for the direct detection of both the oscillation modes of the objects resultant from binary coalescences as well as gravitational echoes and tidal effects\cite{vitor2}, which could be compared with the ones predicted by a perturbative analysis of these models. 

\begin{acknowledgments}
We thank Jos\'{e} P. S. Lemos and Francisco S. N. Lobo for the fruitful discussions.
\end{acknowledgments}



\begin{thebibliography}{99}
\bibitem{darmois}
  G. Darmois, Meml. Sci. Math., {\bf 25}, 565 (1927).

\bibitem{Israel:1966rt} 
W. Israel,  Nuovo Cimento B {\bf 44S10}, 1 (1966).

\bibitem{tolman1}

R. C. Tolman, Phys. Rev. \textbf{55}, 364 (1939).

\bibitem{oppenheimer}

J. R. Oppenheimer and H. Snyder, Phys. Rev. \textbf{56}, 455 (1939).

\bibitem{senovilla1}

F. Fayos, J. M. M. Senovilla, and R. Torres, Phys. Rev. D \textbf{54}, 4862 (1996).

\bibitem{lanczos1}

K. Lanczos, Phys. Z. \textbf{23}, 539 (1922).

\bibitem{lanczos2}

K. Lanczos, Ann. Phys. (Leipzig) \textbf{74}, 518 (1924).

\bibitem{Martinez:1996ni} 
  E.~A.~Martinez,  Phys.\ Rev.\ D {\bf 53}, 7062 (1996).
  
\bibitem{santiago}

S. E. P. Bergliaffa, M. Chiapparini, and L. M. Reyes, Eur. Phys. J. C \textbf{80}, 719 (2020).

\bibitem{Lemos:2017mci} 
  J.~P.~S.~Lemos, M.~Minamitsuji, and O.~B.~Zaslavskii,  Phys.\ Rev.\ D {\bf 95}, 044003 (2017).
  
\bibitem{Lemos:2017aol} 
  J.~P.~S.~Lemos, M.~Minamitsuji, and O.~B.~Zaslavskii,  Phys.\ Rev.\ D {\bf 96}, 084068 (2017).

\bibitem{Lemos:2015ama} 
  J.~P.~S.~Lemos, G.~M.~Quinta, and O.~B.~Zaslavskii,  Phys.\ Lett.\ B {\bf 750}, 306 (2015).

\bibitem{Lemos:2016pyc} 
  J.~P.~S.~Lemos, G.~M.~Quinta, and O.~B.~Zaslavskii,  Phys.\ Rev.\ D {\bf 93}, 084008 (2016).
  
\bibitem{brito}  

R. Brito, V. Cardoso, and J. V. Rocha, Phys. Rev. D \textbf{94}, 024003 (2016).

\bibitem{olea}

J. Crisostomo and R. Olea, Phys. Rev. D \textbf{69}, 104023 (2004).

\bibitem{buchdahl}

H. A. Buchdahl, Phys. Rev. \textbf{116}, 1027 (1959).

\bibitem{rago1}

H. Rago, Astrophys. Space Sci. \textbf{183}, 333 (1991).

\bibitem{dev1}

K. Dev and M. Gleiser, Gen. Relativ. Gravit. \textbf{34}, 1793 (2002).

\bibitem{andreasson1}

H. Andreasson, J. Diff. Eq. \textbf{245}, 2243 (2008).

\bibitem{curiel1}

E. Curiel, Einstein Stud. \textbf{13}, 43 (2017).

\bibitem{farnes1}

J. S. Farnes, Astron. Astrophys. \textbf{620}, A92 (2018).

\bibitem{visser1}

M. Visser and C. Barceló, Cosmo \textbf{99}, 98 (2000).

\bibitem{frauendiener1}

J. Frauendiener, C. Hoenselaers, and W. Konrad, Classical Quantum Gravity \textbf{7}, 585 (1990).  

\bibitem{brady1}

P. R. Brady, J. Louko, and E. Poisson, Phys. Rev. D \textbf{44}, 1891 (1991).

\bibitem{alestas1}

G. Alestas, G. V. Kraniotis, and L. Perivolaropoulos, arXiv:2005.11702.

\bibitem{crawford}

F. S. N. Lobo and P. Crawford, Classical Quantum Gravity \textbf{22}, 4869 (2005).

\bibitem{garcia1}

N. M. Garcia, F. S. N. Lobo, and M. Visser, Phys. Rev. D \textbf{86}, 044026 (2012).

\bibitem{vitorpani}

V. Cardoso and P. Pani, Living Rev. Relativity \textbf{22}, 4 (2019). 

\bibitem{franzin1}

E. Franzin, V. Cardoso, P. Pani, and G. Raposo, J. Physics \textbf{841}, 012035 (2016).

\bibitem{andrea1}

A. Maselli, P. Pani, V. Cardoso, T. Abdelsalhin, L. Gualtieri, and V. Ferrari, Phys. Lett. \textbf{120}, 081101 (2018).

\bibitem{vitor3}

V. Cardoso, S. Hopper, C. F. B. Macedo, C. Palenzuela, and P. Pani, Phys. Rev. D \textbf{94}, 084031 (2016).

\bibitem{vitor2}

V. Cardoso, E. Franzin, and P. Pani, Phys. Rev. Lett. \textbf{116}, 171101 (2016).

\bibitem{garfinkle1}

D. Garfinkle and R. Gregory, Phys. Rev. D \textbf{41}, 1889 (1990).

\bibitem{garfinkle2}

D. Garfinkle and R. Zbikowski, Classical Quantum Gravity \textbf{29}, (2012).

\bibitem{mazur1}

P. O. Mazur and E. Mottola, arxiv:gr-qc/0109035.

\bibitem{binnington1}

T. Binnington and E. Poisson, Phys. Rev. D \textbf{80}, 084018 (2009).

\bibitem{hinderer1}

T. Hinderer, Astrophys. J. 677, 1216 (2008).

\end{thebibliography}
\end{document}